\documentclass[sigconf,nonacm]{acmart}

\AtBeginDocument{
  \providecommand\BibTeX{
    {\normalfont B\kern-0.5em{\scshape i\kern-0.25em b}\kern-0.8em\TeX}
  }
}

\copyrightyear{2025}
\acmYear{2025}
\setcopyright{cc}
\setcctype{by-nc}
\acmConference[Websci '25]{17th ACM Web Science Conference}{May 20--24, 2025}{New Brunswick, NJ, USA}
\acmBooktitle{17th ACM Web Science Conference (Websci '25), May 20--24, 2025, New Brunswick, NJ, USA}
\acmDOI{10.1145/3717867.3717920}
\acmISBN{979-8-4007-1483-2/2025/05}

\usepackage{indentfirst}

\usepackage{graphicx}
\graphicspath{{./images/}} 

\usepackage{subcaption}

\usepackage{array}
\newcolumntype{P}[1]{>{\centering\arraybackslash}p{#1}}

\usepackage{soul}

\makeatletter \g@addto@macro{\UrlBreaks}{\UrlOrds} \makeatother

\begin{document}

\title{GitHub Repository Complexity Leads to Diminished Web Archive Availability}

\author{David Calano}
\orcid{0000-0002-8710-2274}
\affiliation{
  \institution{Old Dominion University\\Department of Computer Science}
  \city{Norfolk, Virginia 23529}
  \country{USA}
}

\author{{Michele C. Weigle}}
\orcid{0000-0002-2787-7166}
\affiliation{
  \institution{Old Dominion University\\Department of Computer Science}
  \city{Norfolk, Virginia 23529}
  \country{USA}
}

\author {{Michael L. Nelson}}
\orcid{0000-0003-3749-8116}
\affiliation{
  \institution{Old Dominion University\\Department of Computer Science}
  \city{Norfolk, Virginia 23529}
  \country{USA}
}

\renewcommand{\shortauthors}{Calano, et al.}

\begin{abstract}
  Software is often developed using versioned controlled software, such as Git, and hosted on centralized Web hosts, such as GitHub and GitLab. These Web hosted software repositories are made available to users in the form of traditional HTML Web pages for each source file and directory, as well as a presentational home page and various descriptive pages. We examined more than 12,000 Web hosted Git repository project home pages, primarily from GitHub, to measure how well their presentational components are preserved in the Internet Archive, as well as the source trees of the collected GitHub repositories to assess the extent to which their source code has been preserved. We found that more than 31\% of the archived repository home pages examined exhibited some form of minor page damage and 1.6\% exhibited major page damage. We also found that of the source trees analyzed, less than 5\% of their source files were archived, on average, with the majority of repositories not having source files saved in the Internet Archive at all. The highest concentration of archived source files available were those linked directly from repositories' home pages at a rate of 14.89\% across all available repositories and sharply dropping off at deeper levels of a repository's directory tree.
\end{abstract}



\maketitle

\section{Introduction}

    Modern software projects are powered by version control software, such as Git,\footnote{\url{https://git-scm.com/}} that enable an inherent level of preservation due to their distributed nature. The Git hosting platform GitHub quickly rose to prominence in 2008 as one of the first centralized Git Web hosts, providing a Web interface to what is otherwise just a collection of files in a directory on a user's local machine.  Web resources in GitHub are a superset of the files in Git, with GitHub providing ``Issues'', ``Discussions'', ``Actions'', and other pages that have no analog in Git itself. 

    An issue with such centralization is that should something happen to an individual Web hosted repository or the entire Web host itself, such as corruption, deletion, or a migration from open to closed source, the hosted project's availability would end for any programmer, academic, or researcher who might need access to it. The only remaining versions available in such cases would be to access preserved versions from various Web and software archiving services. Although dedicated archives, such as Software Heritage,\footnote{\url{https://www.softwareheritage.org/}} exist that mirror and directly preserve publicly available source code, they should not be exclusively relied upon. The dynamic structure of utilizing numerous individual Web pages to represent a repository's source content can often cause problems for traditional Web archiving systems. This characteristic representation often leads to a lack of availability with respect to an archived repository's content and can result in the inability to compile or recreate and utilize software artifacts derived from archived source code. Given the importance of software repositories across the digital ecosystem, we sought to determine how well repositories are being preserved should they no longer be available on the live Web.
    
    To assess the fidelity of archived software repositories, we explore the extent to which hosted software repositories from GitHub,\footnote{\url{https://github.com}} and additionally from various centralized software hosting platforms, such as GitLab,\footnote{\url{https://gitlab.com}} Sourceforge,\footnote{\url{https://sourceforge.net}} and BitBucket,\footnote{\url{https://bitbucket.com}} were archived within the Internet Archive's Wayback Machine. Motivated by our past work \cite{escamilla-icadl23} that measured citations to Git archives in scholarly papers, we now measure how well archived is the archived source tree content. Our goal is to ascertain the accuracy to which software repositories are being archived to the Wayback Machine, as measured by their amount of page damage, as well as to assess the extent to which the source code of Web hosted software repositories are being archived by Web archives. The ultimate aim is to determine if a Web-hosted Git project that has been archived to the Internet Archive's Wayback Machine could successfully be rebuilt using only the archived content. Such a method would be one of the only viable options in the scenario where a project ceases to exist on the live Web and has failed to be preserved in other archives. To achieve these goals, we examined both the presentational pages for archived repositories as well as their archived source trees. Through our examination of archived software repositories, we find that virtually all analyzed source trees archived in the Wayback Machine contain a degree of missing content, leaving them unable to be reconstructed using their archived Web page representations of the source code.

\section{Background}

    \begin{figure}[h]
        \centering
        \includegraphics[width=0.9\linewidth]{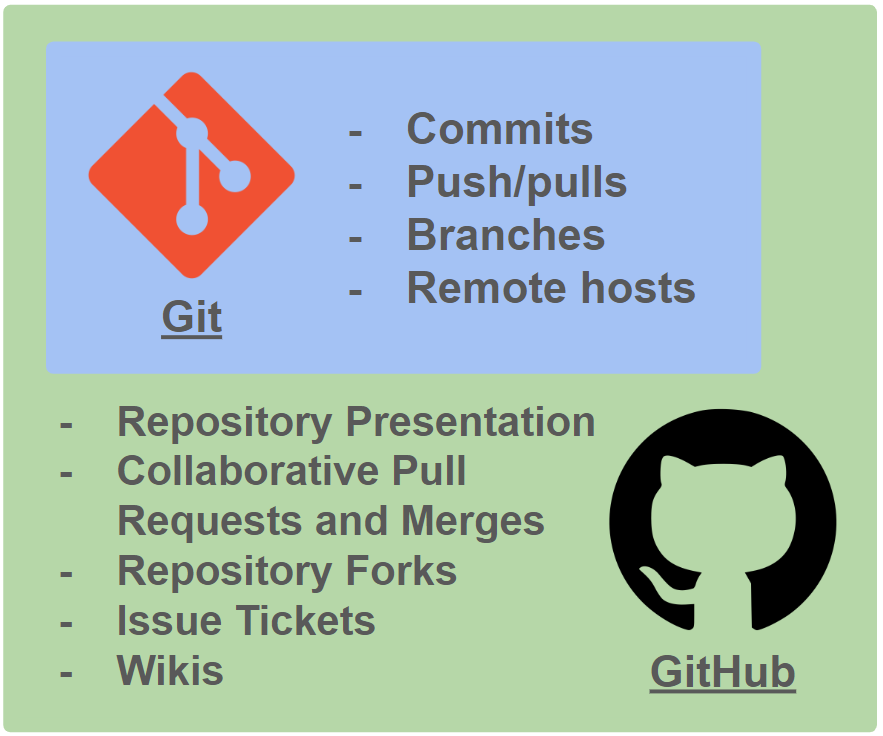}
        \caption{Breakdown of the differences between the version control software Git and the Git hosting platform GitHub.}
        \label{fig:git-vs-github}
    \end{figure}

    Version control software, such as Git, acts as a distributed system for developing software stored as a repository of source files and directories. These distributed Git repositories can be linked to one or more Web servers that act as centralized views of these inherently distributed version control systems that can enhance the set of features provided by Git itself, described in Figure \ref{fig:git-vs-github}. The websites where Git repositories are hosted typically also provide a collection of HTML pages that act as a public portal by which Web users can view and interact with the repository. Numerous websites provide online hosting services for these Git repositories, such as Sourceforge, which was previously popular among software developers as a hub to post source code and packaged binary releases. Shortly after its launch in 2008 GitHub began ramping up in popularity as the dominant platform with its focus on community and social development \cite{how-github-monopolized-code-hosting}. Other alternatives exist as well, such as BitBucket, provided by thMaintaining the Integrity of the Specificationse company Atlassian and integrating with their business-focused development ecosystem, and GitLab, which has also become a community favorite thanks to a rich feature set, open source focus, and the ability for developers to self-host their own versions of the platform.

    \begin{figure*}[h]
        \centering
        \includegraphics[width=0.75\linewidth]{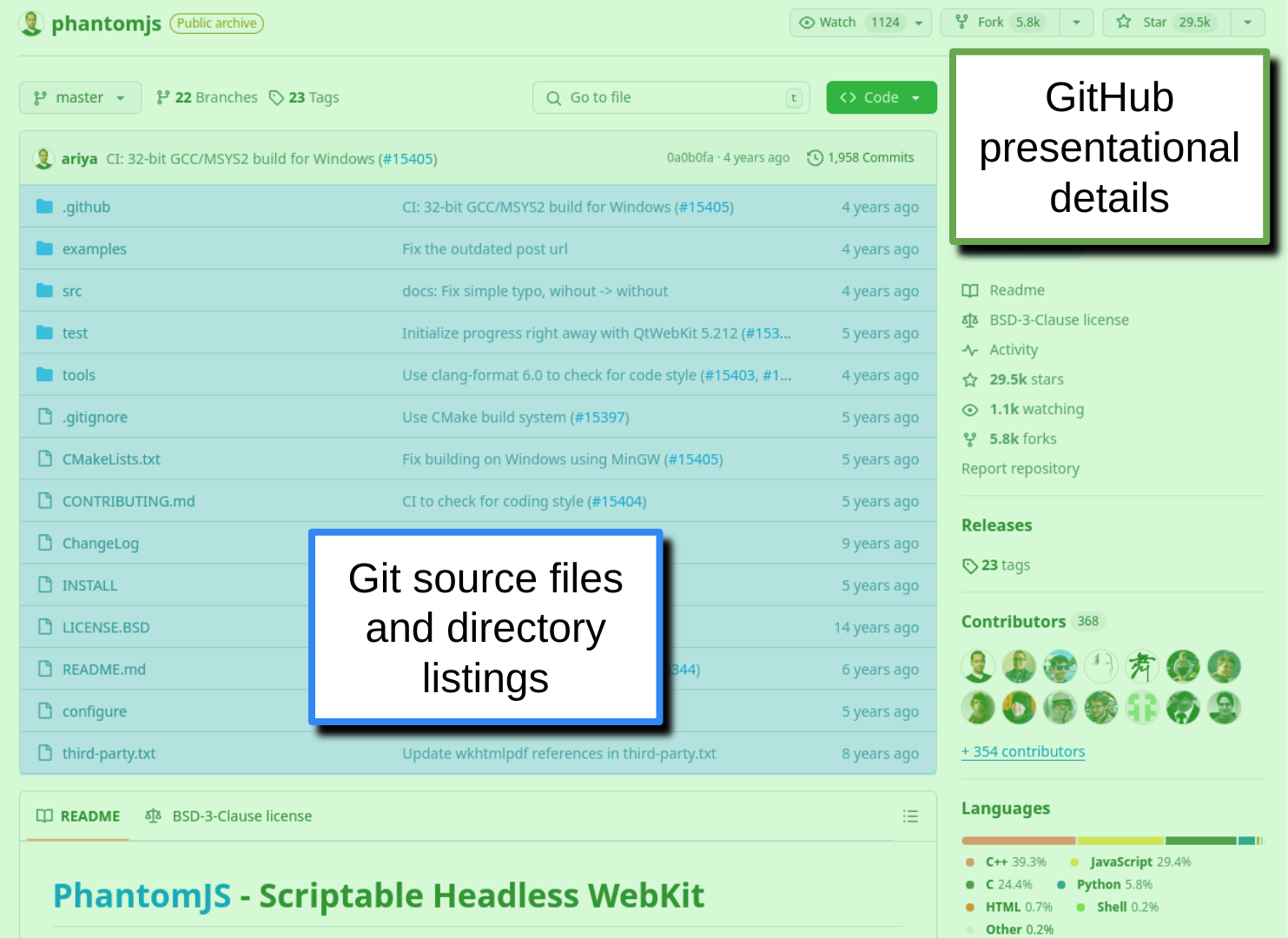}
        \caption{Screenshot of the GitHub home page for the PhantomJS repository showing the embedded representation of the project's top-level source code in blue and the presentational elements of the project generated by GitHub in green.}
        \label{fig:project-homepage}
    \end{figure*}
    
    The software projects hosted on these Web platforms are typically bifurcated into two components, the source files and directories themselves and the related presentational content, such as the rendered README file, issue discussion boards, repository metadata and contributors, and hosted binary release page. This is depicted in Figure \ref{fig:project-homepage}, with the page area highlighted in blue showing the table listing and linking to the repository's top-level source directories and files and the surrounding green area describing the external, associated content generated by GitHub to describe the repository. These presentational pages include the project's landing page rendering a descriptive README and overview of the project, project-specific wiki pages, an index of project release versions, pages detailing historical code tests, and Web pages for issues and pull requests that typically take on the form of discussion boards. Depending upon the Web site where the project is hosted, there may be additional pages detailing the project's code composition, popularity metrics, indexes of contributors and remixed project versions, further non-issue discussions, and other information related to project activity. A repository's README is a special case here, as it is an optional (though generally present) text or Markdown-formatted file that is included within the repository alongside other source files but is almost always rendered into the project's presentational landing page and presented to the user as a traditional HTML Web page. These presentational and descriptive Web pages are not themselves part of the Git repository but are derived from it and then stored separately in another location, such as an external database managed by the Web host. Each source file for a Web hosted repository has two versions, an HTML formatted page that has the source content embedded within it and a raw text version. The HTML page embedded with the source code (example in Figure \ref{fig:phantomjs-file}) is the default page that a user sees when browsing a repository from the Web hosting platform. The raw version is accessed through a button on the source's HTML page and is usually hosted under a different domain or subdomain. For GitHub, the default HTML pages for source files are presented through the normal \url{github.com domain}, while raw versions are served through the \url{raw.githubusercontent.com} domain and exist under a slightly different URI path.

    With software having such a driving importance in our digital lives and increasingly in academic research \cite{software-ac-uk_software}, the preservation of software is a contributor to the issue of academic reproducibility, commonly referenced as a "replication crisis" \cite{10.1145/3170427.3188395, CRL16846, doi:10.1177/0963662520902383}. Open source software and data are important components in cultivating a more open methodology of science to alleviate this crisis 
    \cite{DBLP:journals/corr/abs-1908-05986, 10.1371/journal.pone.0225883, acm-rep-2024-guix-swh, 10.1145/3487553.3524658}, though software is useless if it cannot be found or accessed \cite{10.1371/journal.pone.0115253}. There have been multiple attempts at creating alternative access systems and dedicated archives for Git-based content that are now defunct, such as GHTorrent,\footnote{\url{https://github.com/ghtorrent/}} or no longer updated and maintained, such as the Public Git Archive project \cite{10.1145/3196398.3196464} and GHArchive.\footnote{\url{https://www.gharchive.org/}} In the case of GHTorrent, their provided Web site has now become the home for junkware. Smaller projects and hosts are not the only ones subject to shuttering either, as noted by GitHub's rise as the predominant repository host overtaking the popularity of previous platforms, such as Sourceforge. Another example of this is the closing of the Google Code platform in 2016 \cite{farewell-google-code} that left its over 1.5 million hosted projects no option but to close or migrate to other platforms. While development on this platform is no longer possible, Google has left the code and issues themselves accessible through a static, archived portal. The Google Code platform was once one of the most popular platforms for Web hosted software repositories. The shuttering of such a widely used platform serves to further highlight the fragile ecosystem of software archiving and the importance of supporting and maintaining those still operational, such as GitHub's Archive Program, Software Heritage, and archiving institutions such as the Internet Archive. It is important to have a variety of archives available as the centralization and concentration of archived content within a one institution can cause it to become a single point of failure. Should any centralized institution become the exclusive source of truth for archived content then that content runs the risk of becoming unavailable should that source itself become unavailable.

    Software Heritage is a UNESCO-supported, non-profit initiative aiming to preserve all publicly available source code. Software Heritage focuses specifically on archiving source code and currently holds the largest existing public collection of source code. To date, the archive has preserved the source code for over 300 million software projects collected from 33 major forges, a common term for Git hosting platforms. While smaller, publicly available repository Web hosts might still be overlooked and missed by Software Heritage, they do allow for user requests to archive arbitrary Web repository hosts so they may also be preserved. Despite Software Heritage extensively collecting publicly available source code \cite{10.1145/3379597.3387510}, it does not archive related pages found on software Web hosts, such as the framed HTML pages in which source files are embedded or any project page directly, such as its landing page or issues page, nor does it seek to directly archive binary releases. It is not a direct copy or mirror of Web based Git platforms but collects source code from them as unique digital artifacts for long-term preservation. In contrast, the Internet Archive is more general, primarily preserving Web pages through its Wayback Machine archive, but also containing archives for books, film, audio, and software. The Internet Archive's software archive\footnote{\url{https://archive.org/details/software}} is not equivalent to that of Software Heritage in that while Software Heritage focuses on public source code, the Internet Archive's software archive is oriented towards more packaged releases of digital software code and extracted optical disk images. 

\section{Related Work}

    The interplay between software and scholars has been extensively researched. One of the leading groups on this subject is the Investigating \& Archiving the Scholarly Git Experience project.\footnote{\url{https://investigating-archiving-git.gitlab.io/}} Their research has looked at the importance of preserving software \cite{isage-SAA2020}, the behavior studies of scholars writing code \cite{isage-idcc22-paper}, the usage patterns of scholars working with Git platforms \cite{isage-EnvironScan}, and the challenges faced by scholars and their use of Git \cite{isage-GapAnalysis}. Their work helps to denote how software is utilized and the importance of its preservation outside of software development in academia and library sciences \cite{isage-code4Lib2020, isage-csvConfv5}. This importance goes both ways as well \cite{github-academic-links}, with the development of software often relying on academic research for foundational knowledge, algorithms, and data. With such a strong relationship between software and academic research and our limited ability to trust the content linked from academic publications upon future review, as noted by Jones et al. \cite{scholarly-context-drift}, strong Web archiving systems must be developed, similar to LOCKSS \cite{lockss} and CLOCKSS \cite{clockss}. With the lack of a widespread, distributed network of repository mirrors, this burden heavily rests on efforts by groups such as Software Heritage and the content available in the Internet Archive's Wayback Machine.

    Software Heritage is designed to be a universal archive for software source code \cite{10.1007/978-3-030-52200-1_36, abramatic2018building,dicosmo:hal-01590958, shustek-ieee}. Software Heritage recognizes the aggregated source code from a software repository as important digital objects \cite{cosmo-artifacts}, assigning each collected repository with a unique identifier \cite{dicosmo:hal-01865790} to aid in their handling and use for scientific reproducibility. Their vast collection of archived software source code forms a rich data set that is utilized by researchers in multiple areas of study \cite{pietri:tel-03515795}. This dataset serves as the basis for numerous other research works related to open-source software and data mining \cite{10.1145/3379597.3387512} and as well as helping to further promote an atmosphere of open science \cite{cosmo-software-ecosystems, acm-rep-2024-guix-swh}.

    Despite Software Heritage's extensive catalog of preserved repositories, we have previously found that only 68.39\% of analyzed repositories had been archived there \cite{escamilla-icadl23}. We also explored the expanding prevalence of software repositories cited among scholarly publications from arXiv and PubMed Central \cite{escamilla-tpdl22} and described the tendencies of academic archives to focus on paper content but not associated source code \cite{escamilla-tpdl23}. Our findings showed a rise not only in general URIs referenced among academic papers but a steady increase in the number of references to Web hosted repositories from 2007 to 2021. Across a corpus of more than 11,000 PubMed Central papers, Sourceforge references were originally the most popular, tapering off and ultimately superseded by GitHub as the predominately cited Git hosting platform. For both the PubMed Central corpus and a corpus of over 125,000 arXiv papers, beginning around 2012 the citations to GitHub repositories grew continuously, with repositories hosted on GitHub being cited an average of around 20\% of papers in the arXiv corpus by 2021. Concerning repositories archived in the Wayback Machine, we specifically looked at the presence of repository home pages. We found that despite a growing citation of repositories in scholarly work, around 7.44\% of hosted repository URIs were no longer available on the live Web. Our research in this paper expands this by examining page content and the completeness of archived source code.
    
    The Memento Damage project \cite{deploying-memento-damage} aims to provide an estimate of damage present in an archived Web page. Damage in an archived Web page may be caused by content that was not captured when the original page was archived or issues arising from the archived page's replay that affect the final rendering of the page to the viewer. This includes missing page resources such as image or video elements, page stylesheets, and failed network requests made by JavaScript code responsible for the further loading of content within a page \cite{brunelle-ijdl15b}. The original algorithm calculates a total page damage metric between 0\% and 100\% based on the number of missing page components that are weighted depending on their size, location, and determined importance within the page. This measurement is meant to be performed on archived Web pages, or mementos \cite{rfc7089}, but can be conducted for any Web page. This research was later expanded by looking at issues related to page sub-resources being loaded dynamically by JavaScript \cite{jbrunelle-phdthesis}. Other work peripheral to the Memento Damage project includes that of Kelly et al. \cite{kelly-jcdl14-b} and Banos et al. \cite{banos2016}, which seek to better evaluate the archivability of Web pages, as well as Kiesel et al. \cite{10.1145/3239574}, working to evaluate the quality of archived Web pages through the comparison of archived and live Web pages. These works help to increase the reliability and accuracy of quality and damage measurements for archived Web pages when assessing the extent to which they were archived.

    During our analysis of GitLab and BitBucket pages, we encountered issues related to the use of JavaScript for loading both partial and complete page content. The utilization of JavaScript for page loading can have potentially far-reaching consequences for Web archiving systems. This phenomenon has been found in other research as well, particularly in archived social media pages, as noted by Garg et al. \cite{garg2024challenges} when researching archived Twitter pages and Bragg et al. \cite{bragg-jcdl23} for archived Instagram pages. The impact of JavaScript and its relationship to Web archiving is also detailed by Brunelle et al. \cite{brunelle-archival-crawlers} and Weigle et al. \cite{weigle-jcdl23}.

\section{Methodology}

    To analyze the \textbf{presentational components} of Web hosted repositories, we primarily selected over 10,000 archived public repository pages from GitHub and secondarily 1,000 from GitLab, 600 from BitBucket, and 500 from Sourceforge. The majority of GitHub repository URIs were pulled from data\footnote{https://github.com/oduwsdl/Extract-URLs} that we previously collected \cite{escamilla-tpdl23}. Supplemental URIs were derived from manual searches within each repository host. We used \url{https://github.com/topics/} for GitHub searches, \url{https://sourceforge.net/directory/} for Sourceforge, \url{https://gitlab.com/explore/projects/} for GitLab, and \url{https://bitbucket.org/repo/all/} for BitBucket. To conduct these searches we looked for top and trending repositories within the various platform search pages using keywords such as "education", "data science", "web", and other similar keywords.\footnote{\url{https://gist.github.com/dcalano/15b61a7b0e2054fcf96a94ee6d317651}} In order to assess the page damage of archived repository pages we utilized the Memento Damage tool to visit each URI, retrieve its HTML content, if available, and then assess the condition of the project's home page.

\begin{figure}
        \centering
        \begin{subfigure}[t]{1\linewidth}
            \centering
            \captionsetup{width=.8\linewidth}
            \includegraphics[width=1\linewidth]{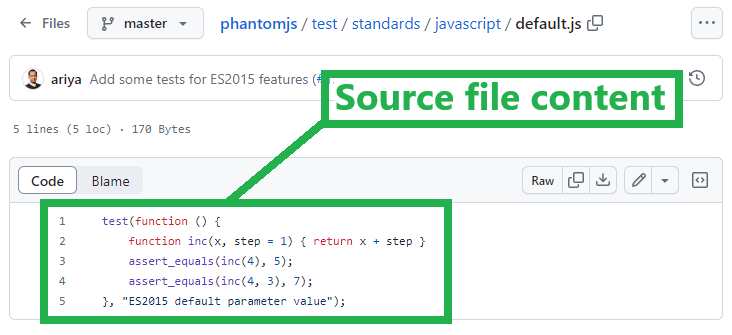}
            \caption{Cropped screenshot of the Web page for a GitHub project source file \url{https://github.com/ariya/phantomjs/blob/master/test/standards/javascript/default.js}.}
            \label{fig:phantomjs-file}
        \end{subfigure}

        \vspace{0.25in}

        \begin{subfigure}[t]{1\linewidth}
            \centering
            \captionsetup{width=.8\linewidth}
            \includegraphics[width=0.9\linewidth]{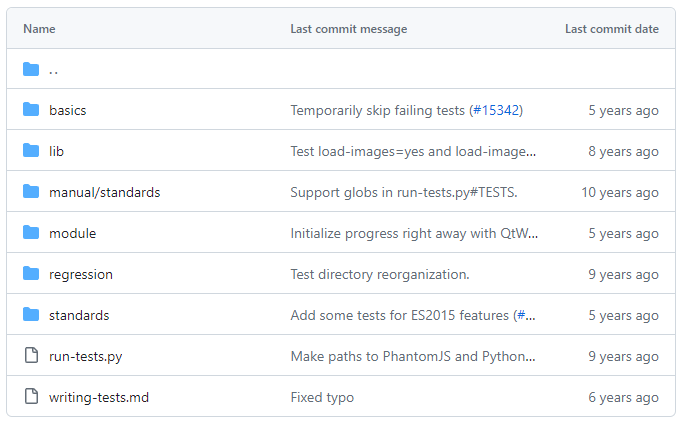}
            \caption{Cropped screenshot of the Web page for a GitHub project source directory, \url{https://github.com/ariya/phantomjs/tree/master/test}, indexing further directories and files.}
            \label{fig:phantomjs-directory}
        \end{subfigure}

        \caption{Cropped screenshots of directory and source file Web pages for GitHub repository.}
        \label{fig:phantomjs-directories-and-files}
    \end{figure}

    The assessment of an archived software project's \textbf{source tree} is a more extensive process, requiring the analysis of numerous pages, and was carried out separately from the presentational analysis. To perform this task we extended the capabilities of the Memento Damage tool with a new crawler component\footnote{\url{https://github.com/oduwsdl/web-memento-damage/tree/code-crawler}} capable of crawling and assessing the state of an archived software project from the Wayback Machine. In this paper, we also report on the archived source trees of hosted software repositories collected from GitHub that we found in the Wayback Machine. We chose to look at archived repositories from GitHub  due to both the overall popularity of the platform and issues stemming from other Web hosts, described in the next section. To determine the extent of a project's archived source tree we first utilized GitHub's API and the Wayback Machine's CDX API\footnote{\url{https://github.com/internetarchive/wayback/tree/master/wayback-cdx-server}} to get a record of the files in the repository's original source tree, their associated Web page representations and raw file URIs, and then their available mementos in the Wayback Machine. For our analysis we targeted the most recent memento with a 200 HTTP status code for each available source file, filtering out older versions for the same source file page, and then conducted page crawls using the Memento Damage tool to examine its content and ensure the page content was properly archived. For these archived source code Web pages, the damage metric is distinct from a traditional page analysis using the Memento Damage tool as any amount of damage to JavaScript, images, stylesheets, or other elements is acceptable, so long as the text for the source file is present on the page. The archived pages that were collected were then separately cross-referenced with the project's original pages from the live Web to determine the completeness of the archived source tree. While we recorded the complete archived source tree, for statistical measurements we chose to utilize the Web pages for source files, shown in Figure \ref{fig:phantomjs-file}, and not those of archived directories, shown in Figure \ref{fig:phantomjs-directory}, within a repository. This is primarily because the content of source files serves as the key element of importance while the Web pages for directories merely serve as indexes linking to other directories and files. Since we can retrieve an overview of available pages from the Wayback Machine's CDX API, we do not need to crawl through directory index pages. In fact, because a project's directory structure can be derived using the source file paths present as part of their URIs, we deemed the directory pages themselves as irrelevant for analysis. While we are able to see the overview of available repository pages in the Wayback Machine, this does not mean we have a complete picture of the archived repository without external comparison to the original source tree. From a view of only the archived content, if a directory page was not archived, then the nested directory pages and source file pages linked within it would also likely be missing and unaccounted for within the Wayback Machine. The exception here would be nested files directly archived through manual submission or other means, circumventing an automated crawl. By analyzing the URI path of such an archived page, we can start to fill in the gaps and make note of a parent directory that is unavailable within the archived source tree, but we would not have visibility of any neighboring files from that missing directory.

\section{Findings}

    Our findings reveal a general lack of completeness among archived repositories. These shortcomings are both due to the size and structuring of Web hosted software projects but also the architecture and traditional methodologies of Web archiving systems. An increased measure of captured page fidelity could be reached through general improvements in existing archival systems. For repository source trees though, the development of new crawlers or improved logic and behavioral patterns specifically designed to address source trees will be necessary to achieve an increase in the completeness of archived content. The practice of Web hosted Git platforms converting a repository source tree into many separate HTML pages is not a good match for Web crawlers or archive systems because, while many pages might be archived, it might only take a single missing source file page to invalidate the underlying source tree's integrity with respect to its ability to be reconstituted. The chance of such a compromise goes up as a function of the number of source files in a repository.

\subsection{Presentation Pages}

\begin{figure}
        \centering
        \begin{subfigure}[t]{1\linewidth}
            \centering
            \captionsetup{width=.8\linewidth}
            \includegraphics[width=0.7\linewidth,height=1.8in]{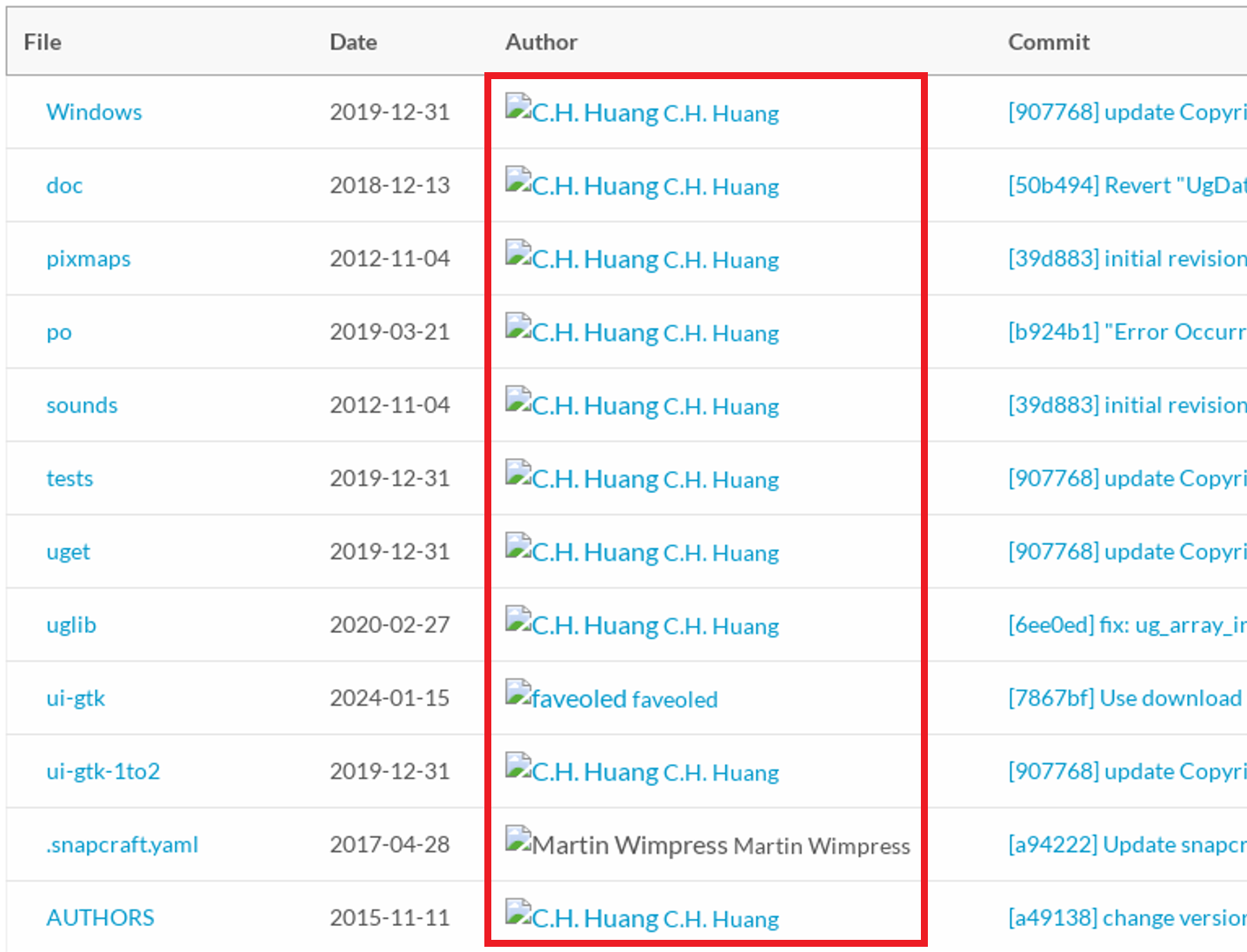}
            \caption{Cropped segment of Sourceforge repository page with 18.29\% damage due to missing commit author avatars, many duplicates for the same author}
            \label{fig:uget2-damage}
        \end{subfigure}

        \vspace{0.25in}
        
        \begin{subfigure}[t]{1\linewidth}
            \centering
            \captionsetup{width=.8\linewidth}
            \includegraphics[width=0.95\linewidth,height=2in]{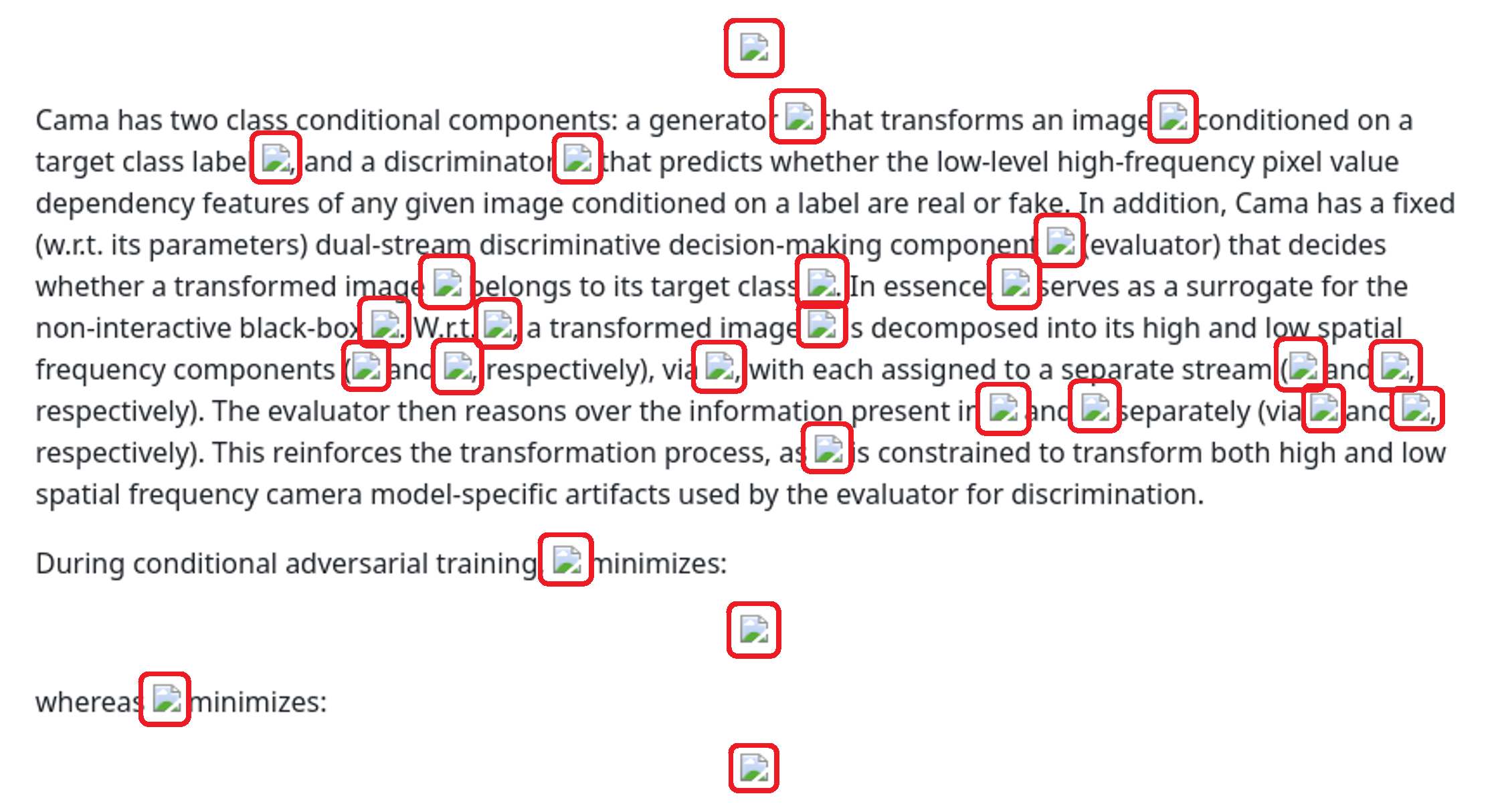}
            \caption{Cropped segment of archived GitHub repository page measured at 49.2\% damage caused by numerous missing full width and inline images scattered throughout repository README}
            \label{fig:cama-damage}
        \end{subfigure}
        \caption{Highlights of missing image elements on two archived repository pages}
        \label{fig:page-damage-comparison}
    \end{figure}

    For the presentational component of our analysis, we found that 31.21\% of GitHub repositories exhibit some form of damage, with around 1.43\% having some form of significant damage to their archived home page when rendered upon playback. Collectively between GitHub and the secondary repository hosting platforms analyzed, around 10.33\% of the sampled repository URIs did not have any mementos in the Wayback Machine. Of those remaining with successful captures, Table \ref{table:repo-damage} shows that 58.22\% exhibited no damage to their home page, 29.85\% exhibited minor damage (defined here as page damage totaling less than 25\%), and 1.6\% exhibited major damage (25\% and above), as measured by the Memento Damage tool. An example of minor damage might be a page\footnote{\url{http://web.archive.org/web/20240211225434/https://sourceforge.net/p/urlget/uget2/ci/master/tree/}} missing a few small or unimportant images, as depicted in Figure \ref{fig:uget2-damage}, while those exhibiting significant damage\footnote{\url{http://web.archive.org/web/20231114065050/https://github.com/jeroneandrews-sony/cama}} had more substantive problems such as large and/or numerous missing images or missing core stylesheets that cause elements to overlap and the page to become unreadable to viewers, as shown in Figure \ref{fig:cama-damage}.

    \begin{table}[h]
        \centering
        \begin{tabular}{|P{0.6cm}||P{1.1cm}|P{1.2cm}|P{1.5cm}|P{0.9cm}|P{0.8cm}|ll}
            \hline
            Host & Sampled & Not Archived & Undamaged & Minor & Major \\
            \hline
            GH & 10,000 & 5.24\% & 63.55\% & 29.78\% & 1.43\% \\
            GL & 1,000 & 4.6\% & 54.5\% & 36.9\% & 4.0\% \\
            BB & 610 & 90.1\% & 0.09\% & 0.002\% & 0\% \\
            SF & 500 & 25.6\% & 18.8\% & 53.6\% & 2\% \\
            Total & 12,110 & 10.33\% & 58.22\% & 29.85\% & 1.6\% \\
            \hline
        \end{tabular}
        \caption{Damage for sampled repositories hosted at GitHub, GitLab, BitBucket, and Sourceforge}
        \label{table:repo-damage}
    \end{table}

\begin{figure}[h]
        \centering
        \includegraphics[width=1\linewidth]{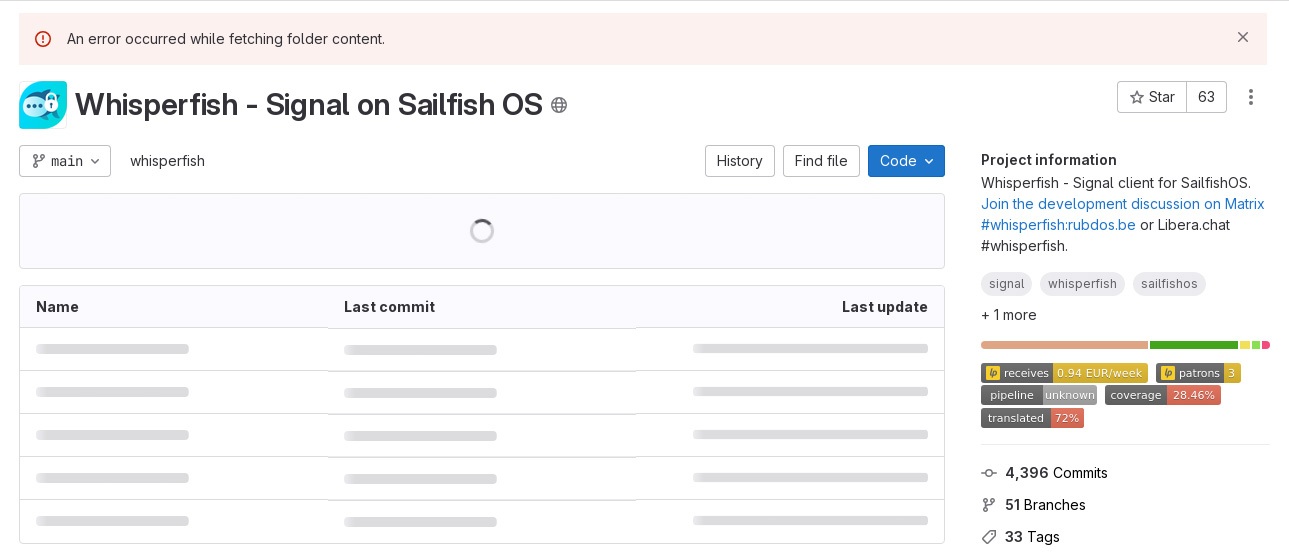}
        \caption{Archived GitLab repository page for WhisperFish project. Only placeholder UI elements were captured instead of the true source table content that is loaded via JavaScript.}
        \label{fig:gitlab-signalfish}
    \end{figure}

    From the set of sampled repository pages, those hosted on GitHub were the most consistently available from the Wayback Machine. Numerous projects on Sourceforge and BitBucket had no mementos available and were only first archived due to the Memento Damage service triggering a crawl action in the Wayback Machine. Of note concerning our findings are those projects hosted through GitLab and BitBucket in which the vast majority are not able to be archived by the Internet Archive beginning around 2020. From what we can deduce, around this time GitLab began to use a more JavaScript-oriented approach that has caused issues with many crawlers, resulting in most projects after 2020 having only the general page skeleton and project overview information available but the ``loading'' placeholder element (gray rectangle) in place of the project's source table, depicted in Figure \ref{fig:gitlab-signalfish}. Similarly, around the same time, it appears that BitBucket began to utilize a Web service called Sentry\footnote{\url{https://sentry.io/}} where the page loads a minimal HTML skeleton and then checks are performed by the Sentry service before performing JavaScript fetch requests to retrieve and load the page information. Each of these affected pages contained an error related to an invalid ``Data Source Name''. For these cases, despite a lack of page content, the Memento Damage service was unable to detect page damage because no network requests were initiated to  be measured for success or failure. Of the pages sampled from BitBucket, only the six archived before 2020 were able to be fully rendered on playback. Many home pages for Sourceforge projects were also originally missing from the Wayback Machine, with more than 100 (of 500) repositories not having been previously archived until a crawl from the Wayback Machine was triggered by the Memento Damage service upon inspection.

\begin{figure}[h]
    \centering
    \frame{\includegraphics[width=.9\linewidth]{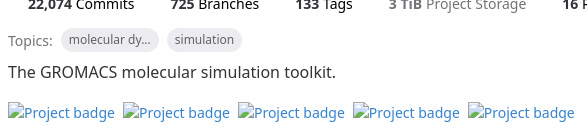}}
    \vskip 0.1in
\frame{\includegraphics[width=1\linewidth]{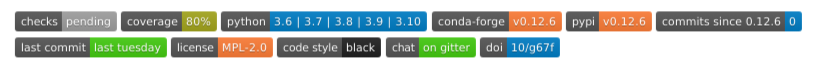}}
    \caption{Example of missing and intact shield badges for Web hosted repository landing pages.}
    \label{fig:badges}
\end{figure}

    For most repository home pages, the largest visual element is the project's README. Much of the page content from the README and the surrounding page itself tended to be static text that is loaded as part of the initial page HTML DOM with damage mostly stemming from missing externally loaded page resources, such as images, embedded GIFs or videos, and stylesheet content. The primary contributors to page damage from missing content were missing user avatars, project logos in the page banner, and images for the dynamic shield badges as depicted in Figure \ref{fig:badges}. Shield badges are small images containing text against a colored background that are generated by an external service and then hot-linked into the project's README using a static URL and often used to show details about a project such as testing code coverage or release version information.

    For the presentational pages of a repository, we find that, as they are largely comprised of text, they retain an overall usefulness for viewers seeking descriptive information about a repository. There might be significant details missing for various projects though, such as images and GIFs used in the project's README to display project screenshots for depicting various aspects or examples of project components.

\subsection{Source Tree}

\begin{figure}[h]
        \centering
        \includegraphics[width=1\linewidth]{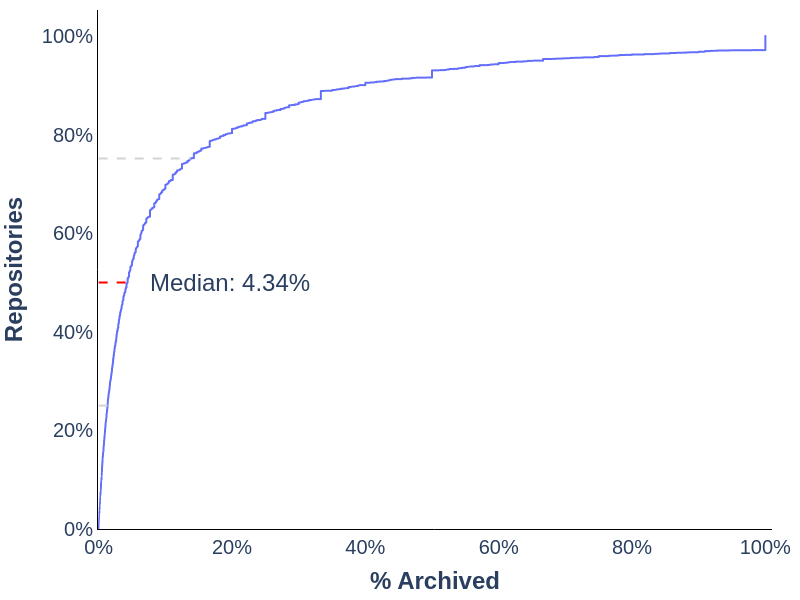}
        \caption{ECDF showing the percentage of repositories with at most $x$ percentage of source files archived, for repositories with a non-zero amount of archived source files.}
        \label{fig:source-ecdf-nonzero}
    \end{figure}

    In analyzing the archived source trees of the collected repositories hosted on GitHub, we found that there were typically only a small number of source files being archived overall. Around 65\% (just over 6,000) did not have any source files archived. Of the remaining repositories that had archived source files, the majority had less than 5\% those source file pages archived within the Wayback Machine, as shown in Figure \ref{fig:source-ecdf-nonzero}, and three-quarters of the repositories had less than 15\% archived. Ultimately, this highlights the lack a general reproducibility for archived software repositories present in the Wayback Machine in terms of being able to rebuild them using their archived source code from Web archives.

\begin{figure*}
        \centering
        \begin{subfigure}{1\linewidth}
            \centering
           \includegraphics[width=0.65\linewidth]{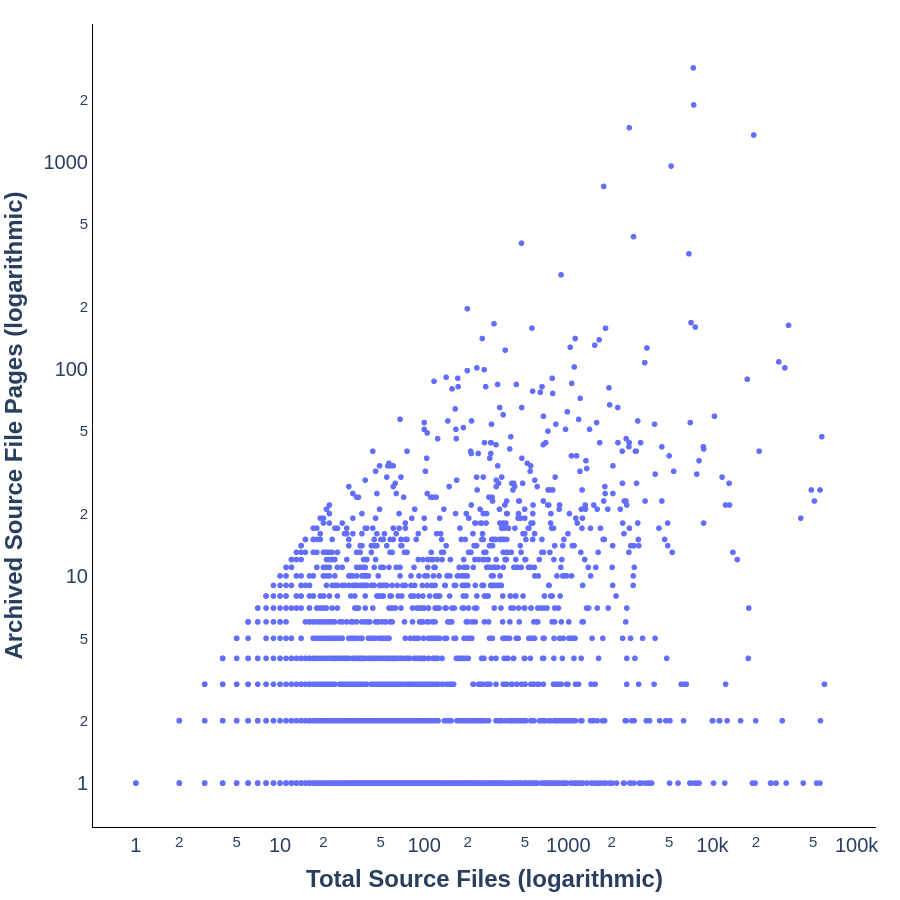}
               \caption{Number of source code Web pages archived vs. total files in the repository (log-log scale)}
               \label{fig:archived-files-count}
        \end{subfigure}

        \vspace{0.15in}
        
        \begin{subfigure}{1\linewidth}
            \centering
            \captionsetup{width=.8\linewidth}
         \includegraphics[width=0.67\linewidth]{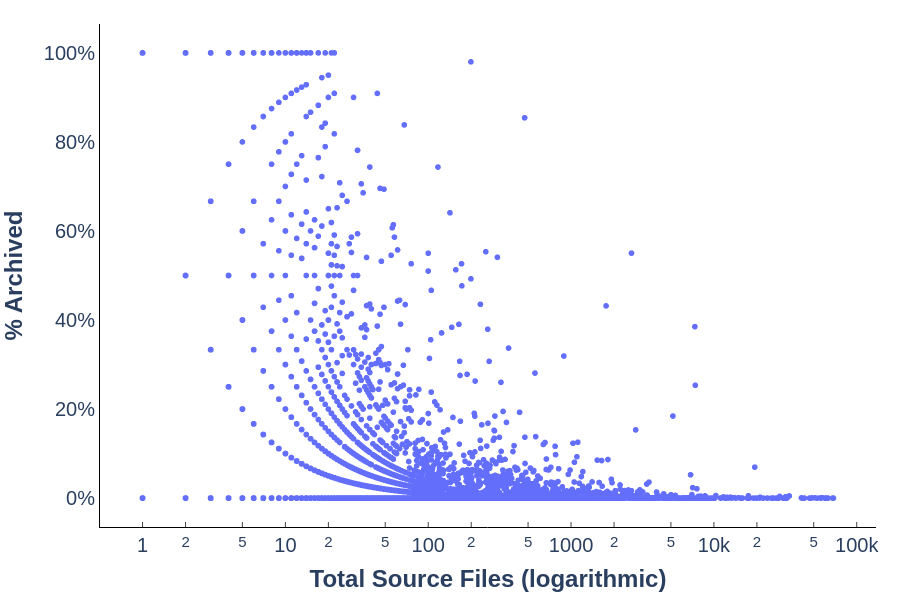}
            \caption{Percentage of source code Web pages archived vs. total files in the repository}
            \label{fig:archived-files-percent}
        \end{subfigure}
        \caption{Archival of source code files in relation to the total number of files in each GitHub repository.}
        \label{fig:archived-files}
    \end{figure*}

     Due to the number of files and depth complexity of source trees, it would appear that more complex source trees generally had lower rates of availability in the Wayback Machine. Those repositories with higher rates of archival had most of their files at or near the source tree root. Only 94 repositories had their source trees completely archived, shown in Figure \ref{fig:archived-files-percent}, with the largest repository among these only containing a total of 22 files. Of those repositories completely archived, only 17 actually contained source code files, while the rest were composed of non-code files, such as Markdown, images, and PDF files, or used as container for a research paper, presentation, or list of Web URIs. When using Spearman's $\rho$ and Kendall's $\tau_b$ metrics to correlate the archived percentage in relation to both file count and structural depth, both relations were found to have very weak correlations, despite the apparent archival trend. For archived percentage in relation to file count, there was a correlation of Spearman's $\rho=0.246~(p=3.325e\text{-}128)$ and Kendall's $\tau_b=0.191~(p=2.825e\text{-}128)$. In correlating repository archival with their structural depths, the correlations were Spearman's $\rho=0.116~(p=2.559e\text{-}29)$ and Kendall's $\tau_b=0.097~(p=1.559e\text{-}29)$. While these statistical correlations are low, we feel that the reasoning behind our findings lies partly in the behavioral mechanisms of the automated crawlers performing the archival of the various repositories. If a Web crawler is retrieving a list of trending repositories from GitHub and then visiting and indexing those pages, there seems to be a low likelihood that it is automatically visiting pages beyond that of the top level home pages gathered. This is indicated by the large number of home pages present from our GitHub sampling in comparison to the number of non-archived source file pages for each repository we examined. With an average file count among all analyzed GitHub repositories of around 505, it would not be impossible for an automated crawler to capture all of the Web pages representing source files, but unless explicitly programmed to recursively traverse all hyperlinks to a repository's source files, some source file pages will inevitably not be captured and the archived source tree would likely not have the ability to be reconstituted naturally.

     Some outlier repositories can be seen (Figure \ref{fig:archived-files-percent}) with a high rate of archival and a not insignificant number of files. These include swift-evolution,\footnote{\url{https://github.com/swiftlang/swift-evolution}} a documentation-oriented repository describing change proposals for the Apple programming language Swift, PowerShell,\footnote{\url{https://github.com/PowerShell/PowerShell}} a popular shell environment by Microsoft, and Konect-Toolbox,\footnote{\url{https:/github.com/kunegis/konect-toolbox}} a network analysis utility for MatLab. For Apple's swift-evolution repository, the bulk of the files are Markdown files linked not to the project's home page but one level lower, nested flatly inside the ``proposals'' directory. While these files might escape typical crawler behavior, this repository has over 10,000 stars, a measure of popularity akin to likes on social media posts, and are all linked to directly from the Swift's main documentation website.\footnote{https://www.swift.org/swift-evolution/} This coupling of in-links from an external Web site, in conjunction with high user popularity driving traffic to these nested pages directly, is an example of how some repositories have higher rates of availability in the Wayback Machine not accounted for by statistical correlation. The PowerShell repository lags quite far behind that of swift-evolution with 55\% of its source files archived, but it also has more than five times the number of total files in addition to a much more complex source tree structure. Although the PowerShell repository does not have numerous external links that help provide inroads directly to repository source files, it does have over four times the number of stars, demonstrating a level of popularity likely to improve its odds of being grabbed by a crawler. The Konect-Toolbox repository is an outlier, falling just shy of a full archival at 98\% for 199 files. This repository is not too complex, having the majority of its files flatly in a subdirectory one level below its root \cite{wsdl-blog-pagination}. The project's repository does not have a significant number of stars, though the project itself has more popularity across various social media websites and through its primary website that may be helping its higher archived percentage. The project's source files have also not changed much, with the latest update being only its README file two years ago. The most recent update behind that was four years ago, with further updates dating up to ten years ago. This largely static source tree and low rate of change are also likely beneficial in allowing more source file pages to become available in the Wayback Machine over time.

\begin{figure}[h!]
        \centering
        \includegraphics[width=1\linewidth]{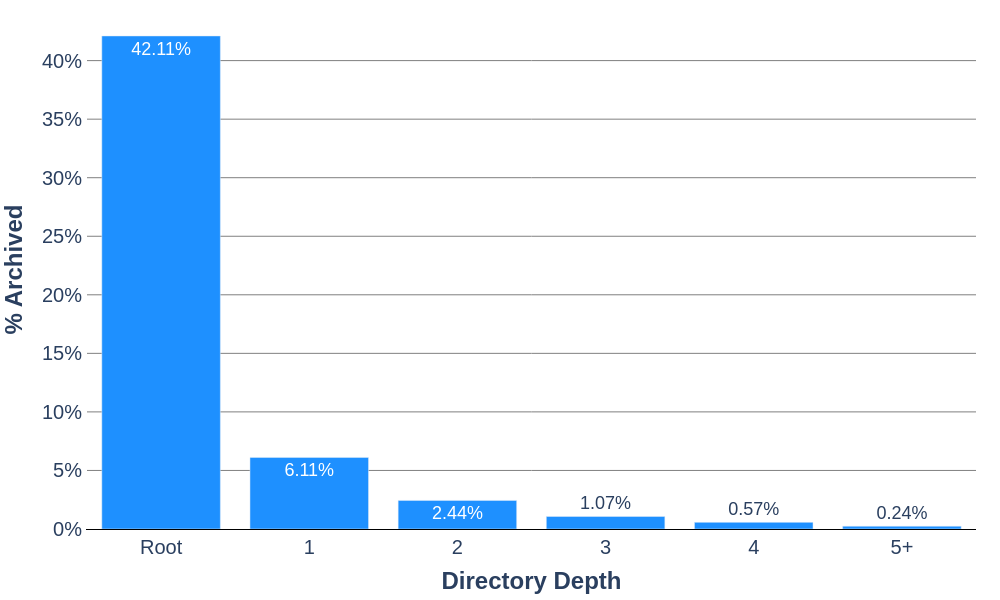}
        \caption{Percentage of source code file pages archived at various directory depths for Github projects with at least one source file archived.}
        \label{fig:source-archived-depth-nonzero}
    \end{figure}

    As Web crawling is a largely automated process and crawlers are not typically set up for depth-wise indexing, their tendency to grab only the content of the direct page they are crawling or content one hop away from it leads to a sharp drop-off in the amount of source files archived beyond the project's root source directory. Those projects that tended to fare the best in terms of the archival completeness of their source tree generally flatter directory structure and a smaller overall file count. This could explain much of the nested file pages of project source trees not being archived in the Wayback Machine, shown in Figure \ref{fig:source-archived-depth-nonzero}. 

    Due to the root project directory being prominently displayed and its source files linked directly from a project's landing page, these source files had the highest archived percentage at just over 42\% for repositories with a non-zero amount of archived source file availability. That rate sharply declined at each subsequent directory depth, being further and further hops away from the landing page, and dropped to just 6.11\% for source directories one level deep, 2.44\% at two levels, and then falling off to just over 1\% and lower at subsequent levels. For our analysis, all directory depths beyond  four were grouped, as there were a diminishing number of repositories utilizing a source tree structure with this level of complexity.
    
\begin{figure}[ht]
        \centering
        \includegraphics[width=1\linewidth]{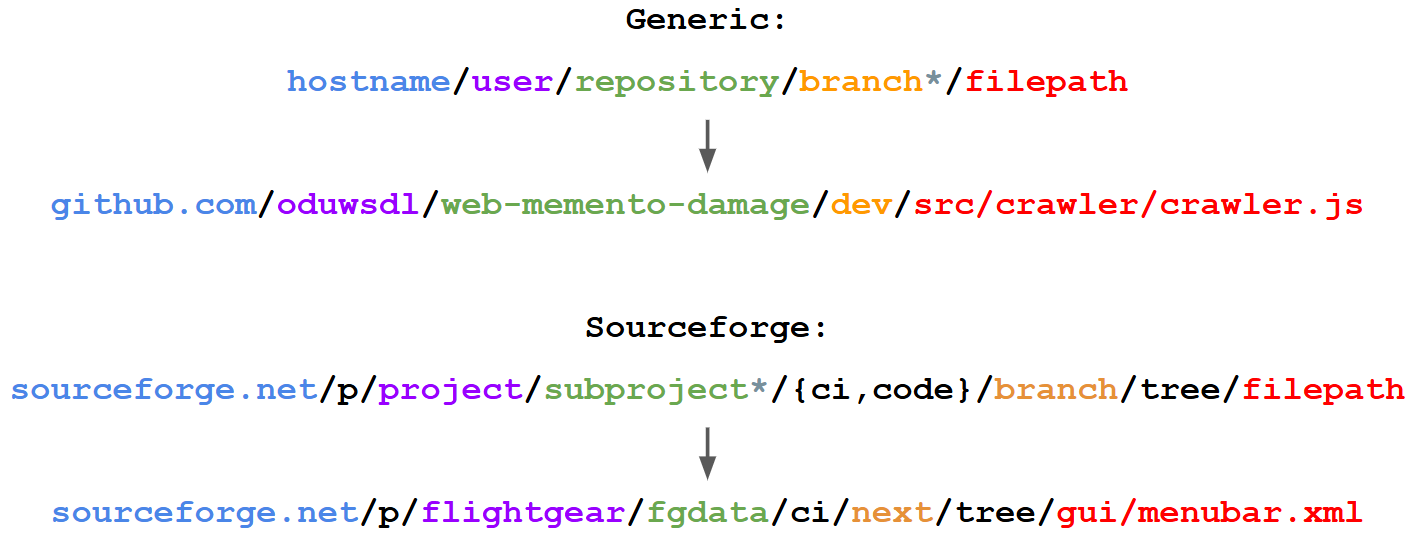}
        \caption{URI schema example for Git hosting platforms. Asterisks indicate sections which may be omitted, such as the branch name in the case of the default branch name being used.}
        \label{fig:url-schemas}
    \end{figure}

    To note, the depth of a repository file or directory is not directly equivalent to the depth of its URI path. Most Web hosts typically use a pattern, shown in Figure \ref{fig:url-schemas}, of hostname/user/repository/type/branch (optional if not default)/filepath, where the owner, repository name, type, and optional branch name form a prefixed path depth in the URI that is additional to the true depth of the source file or directory. Here, type is usually either ``blob'' for directories and ``file'' for files. In most cases, this constant of 3 or 4 can be added when mapping URI and source depths. For the Wayback Machine, this constant would also need to be increased to take into account the path prefixes added to the archived URI-R. Sourceforge is the exception to this rule, which utilizes a more complex schema of \url{sourceforge.net/p/project/subproject/prefix/branch/type/filepath}. Here, the subproject is optional, prefix is either ``code'' or ``ci'', and type is either ``tree'' or ``blob''. In Sourceforge's URI schema, there might not be an organization name, with the source code existing under a single path for the project name and optionally a subpath if subprojects existed. For our analysis and results, we exclusively measured the depth of the source files according to their filepath schema identifier from the URI schema, ignoring the beginning structure that differs among various repository hosting platforms.

\section{Conclusions and Future Work}

    Our analysis of archived Web hosted software repositories highlights the shortcomings of traditional archival methodologies and hosts to faithfully capture the depth of content required to rebuild such projects. We used a combination of API calls for the Internet Archive's Wayback Machine and page crawls using the Memento Damage tool to carry out page analysis for over 10,000 archived repository URIs from GitHub and around 2,000 secondarily from GitLab, BitBucket, and Sourceforge. With regard to the presentational components of hosted repositories, we found that more than 10\% had not been archived and over 30\% had some form of page damage with respect to archived page fidelity. The recent use of JavaScript for the loading of page content on both GitLab and BitBucket also prevented numerous repository pages from being available or usable within the Wayback Machine. An average user without any knowledge of Web archival might see a broken project home page and assume that the entire project is missing. Even if they were knowledgeable about advanced search techniques, such as using the Wayback Machine's CDX API, should there be archived source files for such a project, these pages are likely to suffer the same fate as its home page and be unusable for a viewer. For a user seeking general information about an archived repository on platforms not utilizing JavaScript for page loading, they are most likely to be able to view the text information contained in a project's README when rendered as part of its home page but might not be able to gain useful information should key images be missing from the archived page, such as examples or screenshots. We also measured the extent of source code archival across the collected GitHub repositories. We found that the vast majority of archived repositories could not be rebuilt exclusively from their archived source content within the Wayback Machine. Of the repositories analyzed, an average of 4.72\% of source files were archived. This low average is in part due to only around 35\% of analyzed projects having their source content archived at all. Of those repositories with a non-zero number of archived files, there was still only 13.39\% of each repository archived on average. These source pages are also subject to potential issues due to the usage of JavaScript for loading page content, though the more prevalent issue with the archival of source trees lies in their structure. As each source file is presented using individual HTML pages, archival crawlers must visit each one to completely capture the content of a source tree or risk the archive not having the ability to be reconstructed. As this level of depth is not standard for crawlers archiving the Web, many of the Web pages for deeply nested source code files, which can often be the most crucial or specialized, are not crawled or archived.

    With respect to our analysis of project source trees, we chose to focus on the completeness of archived content. One important factor we did not address was that of temporal coherence with regard to the source files themselves. As software projects have different release versions or milestones over time, the consistency of source file content can change and the archival date becomes a more important factor in determining source validity. Future research will incorporate such temporal analysis as improvements are made to better assess the validity of archived source files for specific software release versions by performing rigorous analysis of changes to the timestamps and commit history of a given software repository. Further improvements to our analysis might be made through static analysis of the code base to improve analytical efficiency and determine source files that are not utilized and could be excluded from the analysis.

    Looking to the future, avenues for further research that could be explored concerning this topic include the expansion of analysis to the complete body of descriptive or presentational pages for repositories, such as wiki, issue, and testing pages. Another interesting avenue of research is the potential to use artificial intelligence systems to fill in the gaps of missing archived source files through code synthesis. For this, machine learning techniques could be harnessed in conjunction with the context provided by other available source files and project information to attempt to generate code files that would enable an incompletely archived repository to be successfully recompiled.


\begin{acks}
This work is supported in part by a grant from the Alfred P. Sloan Foundation. We thank our many colleagues at the Web Science and Digital Libraries research group at Old Dominion University, NYU Division of Libraries, and the Research Library at Los Alamos National Laboratory.
\end{acks}

\bibliographystyle{ACM-Reference-Format}
\bibliography{references}

\end{document}